\documentclass[journal]{IEEEtran}
\IEEEoverridecommandlockouts
\usepackage{amsmath,amsfonts}
\usepackage{algorithmic}
\usepackage{algorithm}
\usepackage{array}
\usepackage{textcomp}
\usepackage{stfloats}
\usepackage{url}
\usepackage{verbatim}
\usepackage{graphicx}
\usepackage{cite}
\usepackage{hyperref}
\usepackage{siunitx}
\usepackage{caption}
\usepackage{subcaption}
\usepackage{todo}
\usepackage{tikz}
\usetikzlibrary{arrows}
\usetikzlibrary{shapes.multipart}
\usetikzlibrary{plotmarks}
\usetikzlibrary{patterns}
\IEEEoverridecommandlockouts
\newcommand\copyrighttext{%
\footnotesize \textcopyright \enspace 2023 IEEE. Personal use of this material is permitted. Permission from IEEE must be obtained for all other uses, in any current or future media, including reprinting/republishing this material for advertising or promotional purposes, creating new collective works, for resale or redistribution to servers or lists, or reuse of any copyrighted component of this work in other works. DOI: \href{https://doi.org/10.1109/LWC.2023.3278181}{10.1109/LWC.2023.3278181}
}
\newcommand\copyrightnotice{%
\begin{tikzpicture}[remember picture,overlay]
\node[anchor=south] at (current page.south) {\fbox{\parbox{\dimexpr\textwidth-\fboxsep-\fboxrule\relax}{\copyrighttext}}};
\end{tikzpicture}%
}

\begin{document}

\title{Analytical Model of 5G V2X Mode 2\\for Sporadic Traffic}

\author{\IEEEauthorblockN{Dmitry Bankov\IEEEauthorrefmark{1}\IEEEauthorrefmark{2}, Evgeny Khorov\IEEEauthorrefmark{1}\IEEEauthorrefmark{2}, Artem Krasilov\IEEEauthorrefmark{1}\IEEEauthorrefmark{2}, Artem Otmakhov\IEEEauthorrefmark{2}\IEEEauthorrefmark{3}}\\
\IEEEauthorblockA{\IEEEauthorrefmark{1}HSE University, Moscow, Russia}\\
\IEEEauthorblockA{\IEEEauthorrefmark{2}Institute for Information Transmission Problems of the Russian Academy of Sciences, Moscow, Russia}\\
\IEEEauthorblockA{\IEEEauthorrefmark{3}Moscow Institute of Physics and Technology, Dolgoprudny, Moscow Region, Russia}
\thanks{The research has been carried out at HSE University and supported by the Russian Science Foundation (Grant No 21-79-10158, https://rscf.ru/en/project/21-79-10158/)}
}

\maketitle
\copyrightnotice

\begin{abstract}
	5G Vehicle-to-Everything (V2X) is a promising technology to satisfy the increasing demands of intelligent transportation systems.
	Emerging V2X applications with a high level of automation impose very strict requirements on latency (less than 10 ms) and reliability (higher than 99.99\%).
	For sporadic traffic, such demands can be satisfied with a distributed channel access method called Mode 2. 
	The letter proposes an analytical model of Mode 2 that estimates the packet loss rate and the network capacity taking into account the peculiarities of Mode 2 and --- in contrast to the existing models --- provides the accuracy required in the emerging V2X scenarios. The model can be used to find the optimal transmission parameters that maximize the network capacity and/or to select the required bandwidth.
\end{abstract}


\section{Introduction}
\label{sec:intro}
Successful development of intelligent transport systems and autonomous vehicles relies on the availability of standardized wireless technologies that can provide reliable communication: (i) between vehicles and (ii) between vehicles and infrastructure. 
To satisfy such a demand, 3GPP develops the 5G Vehicle-to-Everything (V2X) technology~\cite{v2x-tutorial}. 
In~\cite{3gpp-v2x-scenarios}, 3GPP defines typical V2X scenarios and the corresponding quality of service (QoS) requirements. The specific values of the requirements depend on the scenario and the level of automation (LoA). For example, the sensor information sharing scenario with a low LoA requires end-to-end latency below 100~ms and reliability above 99\%. 
In contrast, cooperative lane sharing and other emerging scenarios with high LoA require latency below 10~ms and reliability above 99.99\%. 

Depending on the application, vehicles (called UEs in 3GPP specifications) can generate periodic or sporadic traffic. In particular, a UE can periodically broadcast Cooperative Awareness Messages (CAMs) describing its current state: position, velocity, etc.
When a UE detects some dangerous situation on the road which occurs at a random time moment, it broadcasts a Decentralized Environmental Notification Message (DENM).

To transmit a CAM/DENM, a UE shall access the channel. For that, 5G V2X introduces two methods: (i) centralized resource allocation by a base station (BS) called Mode 1, (ii) distributed resource allocation by UEs called Mode 2.
To obtain a resource with Mode 1, the UE sends a request to a BS. Then the BS allocates a  resource from the common pool and sends the grant to the UE. With a single grant, the BS can allocate periodic resources to efficiently serve periodic traffic.    
In contrast, with Mode 2, which is a sort of multichannel slotted ALOHA, a UE randomly selects a slot and a continuous range of subchannels not yet reserved by other UEs. The required number of subchannels depends on the packet size and used modulation and coding scheme (MCS).  
For both Modes, a UE can reserve resources in the subsequent slots to perform retransmissions, which improves reliability.

In the letter, we consider an urban scenario with UEs connected to BSs. The CAM traffic is served with Mode 1. The DENM traffic is served with Mode 2 because: (i) requesting resource for each randomly appearing packet induces high overhead, (ii) for latency-critical applications the request/grant procedure may exceed the packet delay budget. Also, we assume that separate resource pools are allocated for CAM and DENM traffic. As shown in~\cite{slicing}, such a slicing approach avoids interference between CAMs and DENMs and improves the reliability. An open question is how to dynamically select the bandwidth allocated for each slice depending on the traffic load and its QoS requirements. Alternatively, for a given bandwidth, what is the \textit{network capacity}, i.e., the maximal load for which the QoS requirements are satisfied? In the letter, we address these questions for the DENM slice.       

Mode 2 and the similar channel access method used in LTE V2X systems are studied experimentally~\cite{mikami2020field} or with simulations~\cite{molina2017lte, romeo2021supporting}.
However, in the case of high reliability (e.g., 99.99\%), the evaluation of very low packet loss rate (PLR) requires rather long experiments or simulations with hundreds of thousands of packets to catch rare losses and get the necessary statistics. Thus, neither experiments nor simulations can estimate the capacity in real time, which is needed for optimal network configuration. An alternative is analytical models that quickly estimate the network capacity and PLR for given scenario parameters.   
Several models of slotted multi-channel ALOHA  have been proposed in the literature, e.g.,~\cite{baccelli2006aloha, nba2020discrete, vinel2019cellularV2X}, where \cite{vinel2019cellularV2X} considers channel access method used in LTE V2X systems. However, these studies do not consider many peculiarities of 5G V2X Mode 2 and, thus, the error is too high as we show in Section~\ref{sec:results}. Specifically, the cited models do not take into account the fact that a UE cannot simultaneously transmit and receive signals in the same slot (a.k.a. the \textit{half-duplex} problem).
Also, they assume that packets of different UEs can overlap only entirely in the frequency domain, while  Mode 2 allows partial overlapping.
Finally, these models do not consider the correlation between the collisions of successive transmission attempts.

Our letter presents an analytical model of DENM transmission with 5G V2X Mode 2 without the mentioned drawbacks.
The model accurately estimates PLR as low as $10^{-5}$, which is the target for emerging scenarios. The model can be used to configure Mode 2 (e.g., the number of retransmissions) in order to maximize the network capacity in a given scenario.

The rest of the letter is structured as follows.
Section \ref{sec:scenario} describes the considered scenario and the problem statement.
Section \ref{sec:model} develops the analytical model.
Section \ref{sec:results} discusses the numerical results.
Section \ref{sec:conclusion} concludes the letter.

\section{Scenario and Problem Statement}
\label{sec:scenario}

Consider a scenario with $N$ ($N\gg1$) UEs located on a straight road so that the distance between neighboring UEs has an exponential distribution with parameter $\phi$. 
Each UE broadcasts DENM packets within its communication range $R$ ($R \gg \phi^{-1}$). 
The path loss at distance $r$ is given by the function $l(r) =(Ar)^{-\beta}$, where $A>0$ and $\beta\geq 2$.
Having finished delivering the previous packet (either successfully or not), each UE generates a new one after a random time with the exponential distribution with parameter $\lambda$.

To transmit data, the UEs use Mode 2. The time is divided into slots of duration $\tau$.
Within each slot, $B$ subchannels are available for transmission, where $B$ depends on the bandwidth allocated for the DENM slice.
When a packet arrives, the UE randomly selects a range of $M$  out of $B$ subchannels for packet transmission in the next slot.
While in reality $M$ depends on the packet size and MCS, we assume that for all UEs the size of packets, MCS, and the total transmit power $S$ are the same. 
If packets overlap in time and frequency domains, they may be lost. We model losses with a widely-used exponential effective SINR mapping (EESM) model~\cite{lagen2020new}. Specifically, we compute the signal-to-interference-and-noise ratio (SINR) for each subchannel where the packet is transmitted. The packet is received successfully if the value of EESM exceeds the threshold $T$, which depends on the MCS.

To improve reliability, the UE makes $\nu$ repetitions. Because of the delay budget $D^{QoS}$, the slots for repetitions shall be selected within the window of $W - 1$ slots, where $W=\lfloor \frac{D^{QoS}}{\tau} \rfloor$. We assume that UEs do not use the resource reservation mechanism for repetitions (a UE transmitting a packet can notify the neighbors about future slots and subchannels that will be used for repetitions).  The analysis of this mechanism is the direction for our future work. As UEs are not aware of the future transmission of each other, to avoid correlated errors, the UEs equiprobably select $\nu$ slots within the window $[1,W-1]$ for the corresponding repetitions. For each repetition, UE randomly selects $M$ out of $B$ subchannels.       

As the delay budget  $D^{QoS}$ is small, for typical vehicle speeds the inter-UE distances do not change significantly and the UEs can be considered static within $D^{QoS}$.

For this scenario, we develop a model that finds the probability $PLR(\lambda)$ that a UE within the transmitter's communication range does not receive the packet as the function of the load $\lambda$ and other parameters.
With this function, we obtain the network capacity as $C = \max\left\{\lambda: PLR(\lambda) \leq PLR^{QoS}\right\}$, where $PLR^{QoS}$ is the PLR constraint.

\section{Analytical Model}
\label{sec:model}
\subsection{General idea}
\label{sec:model_general}
The model has two main assumptions.

According to the scenario, the number of UEs in the network is large, and the distance between neighbors has an exponential distribution.
Therefore, the \emph{first assumption} of the model is that such a system is approximated by a Poisson Point Process (PPP) on a straight line with the parameter $\phi$.

On average, a UE waits for $\frac{1}{\lambda\tau}$ slots before generating a packet and makes $\nu$ repetitions in the next $W \frac{\nu}{\nu + 1}$ slots. 
Thus, the UE's transmit probability in a given slot equals
\begin{equation}
	\label{eq:p}
	p = \frac{1 + \nu}{\frac{1}{\lambda \tau} + W \frac{\nu}{\nu + 1}}.
\end{equation}
Our \emph{second assumption} is low traffic intensity, i.e., $p \ll 1$.

Let us randomly select a UE (transmitter, TX) sending a packet to another UE (receiver, RX) located within the range $R$.
According to the property of PPP, the distance $r$ between them is distributed uniformly, so we find $PLR$ as
\begin{equation*}
	\label{eq:plr}
	PLR = \int_0^R PLR_r(r) \frac{dr}{R},
\end{equation*}
where $PLR_r(r)$ is PLR for distance $r$ between TX and RX.

TX's transmission attempts can fail due to several reasons.
First, it can fail because the RX is transmitting its packet itself and cannot simultaneously receive a packet.
Second, if RX is not transmitting a packet, TX's transmission can fail due to a collision with some other UE's transmission.

When TX makes its first transmission attempt, we have no information about the other UEs' transmissions, so each UE makes a transmission attempt with probability $p$.
However, if this transmission attempt fails, we know that a UE (further related to as \emph{active} UE) has made a transmission attempt and will probably make repetitions in the next slots which will increase the probability of a new collision because it is a priori close enough to RX to make a collision and because repetitions are made within a short time window.
Thus we need to statistically distinguish the ``mean'' flow of UEs' transmissions and the flow of active UEs' transmissions.

With this idea in mind, to find $PLR_r(r)$, we introduce a function $V_{t, c}$, which equals the probability of not delivering the packet within $t$ transmission attempts when the number of active UEs equals $c$. Note that the required $PLR_r(r) = V_{\nu + 1, 0}$.
For $t = 0$, $V_{t, c}$ equals $1$, i.e., the packet cannot be delivered without any attempts, otherwise:
\begin{equation}
	\label{eq:vtc}
	V_{t, c} = p V_{t - 1, c} + \left(1 - p\right) \left(U_{t, c} + Y_{t, c}\right), \quad t > 0, 
\end{equation}
The first summand describes a case when TX makes a transmission, but RX cannot receive the packet because it is transmitting a packet itself (with probability $p$, specified in~\eqref{eq:p}).
The second summand describes a case when RX does not transmit (with the probability $1 - p$).
In this case, the packet can be lost because of a collision with at least one transmission of the ``mean'' flow of UEs' transmissions (term $U_{t, c}$), or because of a collision with an active UE's transmission (term $Y_{t, c}$).
We derive $U_{t, c}$ and $Y_{t, c}$ below.

\subsection{Mean flow of UE transmissions}
Consider a transmission attempt, assuming that other UEs make their attempts independently with probability $p$.
The packet is decoded successfully when SINR in the subchannels used for packet transmission is greater than  $T$, which happens with the probability $P_{s}$.
We assume that the packet may not be received if the interfering UEs are at a distance not exceeding $\bar{r}$.
Then the success probability equals
\begin{equation}
	\label{eq:p_success_first}
	P_{s} (r) = \sum\limits_{n=0}^{\infty} H_n  \sum\limits_{k=0}^n G_{n, k}(p) \cdot Q_{k} (r),
\end{equation}
where $H_n$ is the probability that $n$ UEs besides TX are closer than $\bar{r}$ to RX obtained from the PPP properties,
$G_{n, k}(p)$ is the probability that $k$ UEs out of $n$ transmit their packets in the considered slot, which follows the binomial distribution:
$	G_{n, k}(p) = {n \choose k} p^k  (1-p)^{n-k},
$ 
 and $Q_{k}$ is the probability of successful reception of the TX's packet when $k$ other UEs transmit simultaneously, which is found as follows.

Let $P_{M, m}$ be the probability that two packets occupying $M$ subchannels overlap in $m$ subchannels, and $q_{M, m}$ be the corresponding probability of successful packet reception. Then
\begin{equation}
	\label{eq:Q_1}
	Q_1 = \sum\limits_{m=0}^M P_{M, m} \cdot q_{M, m}.
\end{equation}
We find $P_{M, m}$ by solving a simple combinatorial problem:
\begin{equation*}
	\label{eq:Pijk}
	P_{M, m} = \begin{cases}
		0, & m < 2 M - B,\\
		\frac{(B + 2 - 2M) \cdot (B + 1 - 2M)}{(B + 1 - M)^2}, & m = 0, 2M < B,\\
		\frac{2(B + m + 1 - 2M)}{(B + 1 - M)^2}, & 0 < m < M, \\
		\frac{1}{B + 1 - M}, & m = M.
	\end{cases}
\end{equation*}

Let us find $q_{M, m}$ in~\eqref{eq:Q_1}.
For successful packet delivery, EESM in the subchannels where the packet is transmitted shall exceed the threshold $T$.
SINR in subchannels without interference equals $SINR_0 =\frac{l(r) S}{M\sigma}$, where $\sigma$ is the thermal noise power in one subchannel.
In subchannels with interference, SINR equals $SINR_1 =\frac{l(r) S}{l(r_{int}) S + M\sigma}$, where $r_{int}$ is the distance between RX and an interfering UE.

According to the EESM model~\cite{lagen2020new}, the effective SINR for two packets of width $M$ overlapping in $m$ subchannels equals
\begin{equation*}
	\label{geometric_averaging}
	EESM = -\gamma \ln \left(\frac{M-m}{M} e^{-\frac{l(r) S}{\gamma M \sigma}} + \frac{m}{M} e^{-\frac{l(r) S}{\gamma (l(r_{int}) S + M \sigma)}}\right),
\end{equation*}
where $\gamma$ is an approximation constant.
If we denote
\begin{equation*}
	\label{xi}
	\xi_{M, m} (r) = \frac{M}{m} e^{-\frac{T}{\gamma}} - \left(\frac{M}{m} - 1 \right) e^{-\frac{SINR_0}{\gamma}},
\end{equation*}
then $EESM > T$ means that $r_{int}$ should be greater than $0$ for $\xi \geq 1$, and otherwise greater than
\begin{equation*}
	\label{eq:r_min}
	\rho_{M, m} (r) = \frac{1}{A}\left[-\frac{(Ar)^{-\beta}}{\gamma \ln (\xi_{M, m}(r))} - \frac{\sigma M}{S}\right]^{-\frac{1}{\beta}},
\end{equation*}

By the property of PPP, $r_{int}$ is distributed uniformly over $[0, \bar{r}]$.
Thus, the probability that the interfering device is sufficiently far from RX equals
\begin{equation}
	\label{eq:q_Mm}
	q_{M, m} = \mathcal{P} \left(r_{int} > \rho_{M, m}\right) = 1 - \frac{\rho_{M, m} (r)}{\bar{r}}.
\end{equation}

We assume that the interference from other UEs can be considered independently for each UE: $Q_{k} = Q_1^k$, substitute \eqref{eq:Q_1} and \eqref{eq:q_Mm} in \eqref{eq:p_success_first}, and obtain 
\begin{equation}
	\begin{split}
		\label{eq:p_success_final}
		P_{s}\left(r\right) &= e^{-2\phi \bar{r}} \sum\limits_{n=0}^{\infty} \frac{(2\phi \bar{r})^n}{n!} \sum\limits_{k=0}^n {n \choose k} (Q_1 p)^k (1-p)^{n-k}\\
		&= e^{-2\phi \bar{r} (1-Q_1)p} = e^{-2 \phi p \sum_{m = 1}^{M} P_{M, m} \rho_{M, m} (r)}.
	\end{split}
\end{equation}
Note that the dependency on $\bar{r}$ has been canceled.

\subsection{Transmissions by active UEs}
\label{sec:active-ue}

An active UE makes $\nu$ repetitions during $W - 1$ slots, so its transmission probability in a slot equals $P_{r} = \frac{\nu}{W - 1}$.

Let an active UE and TX make a collision at the first transmission attempt and then make a repetition in the same slot.
When finding the repetition success probability $P_{nc}$, we  consider that because of the prior collision, the UE and TX are located at such a distance from RX that a collision between them is possible for some overlap between their packets:
\begin{equation*}
	\label{P_int_first}
	P_{nc} = 1 - \frac{\sum\limits_{m=1}^M \sum\limits_{l=1}^M P_{M,m} P_{M,l} \cdot \min \left(\rho_{M, m}(r), \rho_{M, l}(r)\right)}{\sum\limits_{m=1}^M P_{M,m} \cdot \rho_{M, m}(r)},
\end{equation*}
where the numerator is the joint probability of failure during the first attempt and repetition, and the denominator is the failure probability of the first attempt.

Let us find $Y_{t, c}$ used in \eqref{eq:vtc}, which is the probability that a packet collides with a packet from $c$ active UEs and with no packet of the ``mean flow''.
An active UE makes a repetition with probability $P_r$, and the number $i$ of such UEs has a binomial distribution $G_{c, i}(P_r)$.
When $i$ UEs make a repetition simultaneously with TX, a collision happens with probability $1 - P_{nc}^i$.
Also, if an active UE makes its last repetition, it stops being active, which happens with probability $\frac{1}{\nu + 1}$, and the number $j$ of such UEs has a binomial distribution.
Thus:
\begin{equation*}
	\resizebox{\linewidth}{!}{$Y_{t, c} = P_s \sum\limits_{i = 1}^{c} G_{c, i}(P_r) \left(1 - P_{nc}^i\right) \sum\limits_{j = 0}^{i} G_{i, j}\left(\frac{1}{\nu + 1}\right) V_{t-1, c - j}$.}
\end{equation*}

In case of a collision with the ``mean'' flow of UE transmissions described by $U_{t, c}$ in~\eqref{eq:vtc}, again, $i$ out of $c$ active UEs can make a transmission attempt, and $j$ out of such UEs make the last attempt.
Also, we consider that $k$ UEs become active after such a collision and estimate the probability of such an event, provided that the collision has happened, as $p^k$.
Thus
\begin{equation*}
	\resizebox{\linewidth}{!}{$U_{t, c} = \left(1 - P_s\right) \sum\limits_{i = 0}^{c} G_{c, i}(P_r) \sum\limits_{j = 0}^{i} G_{i, j}\left(\frac{1}{\nu + 1}\right) \sum\limits_{k = 1}^{\infty} p^{k - 1}  V_{t-1, c + k - j}$.}
\end{equation*}
Note that in the inner sum, the summands decrease exponentially, so to calculate this formula, it is sufficient to take only the first $\lceil \log_p{PLR^{QoS}} \rceil$ summands.

\section{Results and Discussion}
\label{sec:results}

To validate the developed model, we use the NR V2X module~\cite{ns-3-v2x} of NS-3 simulator with optimized channel modeling to speed up simulations.
Unlike the analytical model, the simulation does not use most assumptions stated in Section \ref{sec:model}, e.g., it considers a finite number of UEs instead of a PPP, uses a $BLER(SINR)$ curve to find the probability of successful reception of a packet, instead of comparing the SINR with a threshold $T$, and does not assume that the interference of UEs is independent of the other UEs (i.e., $Q_k \neq Q_1^k$).

We consider a scenario with $N = 1000$ UEs, which, as in \cite{romeo2021supporting}, generate 300-byte packets and use MCS 6 to broadcast them within radius $R = 200$~m.
Each packet occupies $M = 3$ subchannels.
The slot duration is $\tau=0.5$~ms.
The latency requirement is $D^{QoS}=10$~ms.
UEs' transmission power is $S=23$~dBm.
The pathloss parameters are $A=36$~m$^{-1}$, $\beta=3$.
The channel bandwidth is $B = 10$ subchannels.
In the analytical model, SINR threshold $T$ is $2.3$~dB (corresponds to the block error rate of 0.5 in NS-3) and $\gamma=1.15$.

\begin{figure}
	\centering
	\includegraphics[width=0.9\linewidth]{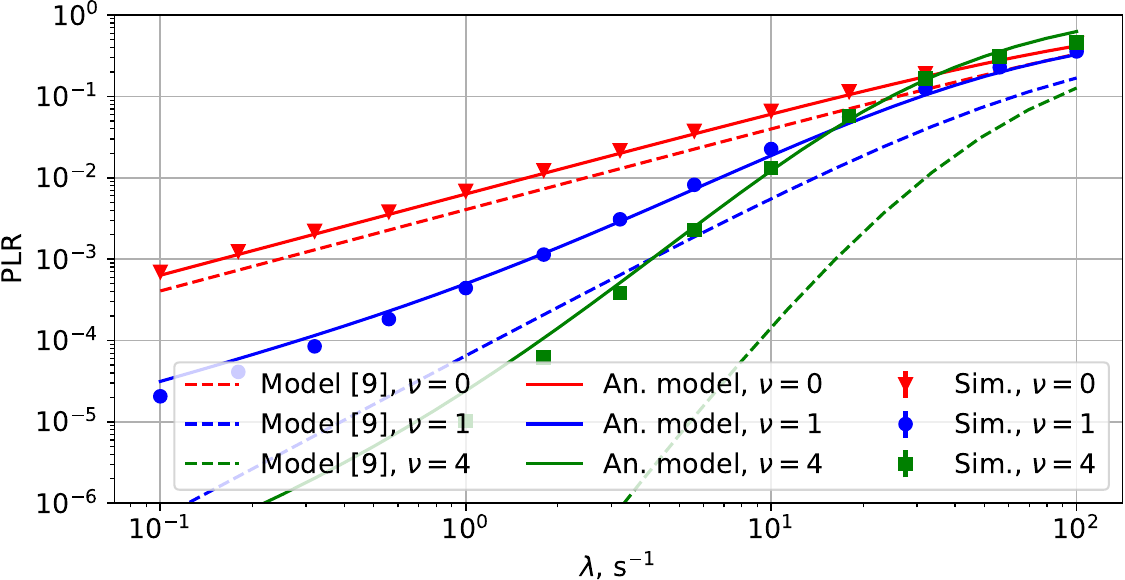}
	\caption{Model validation.}
	\label{fig:validation}
\end{figure}
\begin{figure}
	\centering
	\begin{subfigure}{0.45\linewidth}
		\includegraphics[width=0.96\linewidth]{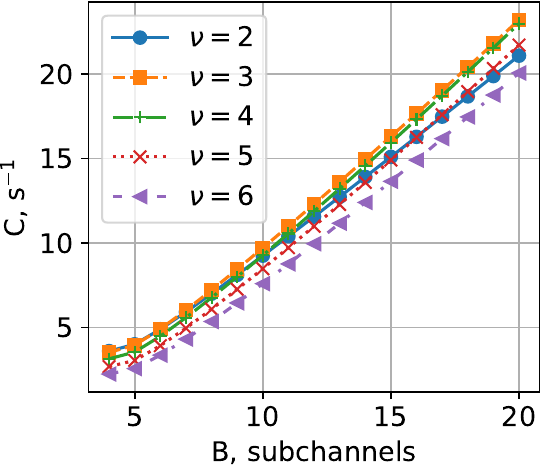}
		\caption{$PLR^{QoS} = 10^{-2}$.}
	\end{subfigure}
	\begin{subfigure}{0.45\linewidth}
		\includegraphics[width=0.96\linewidth]{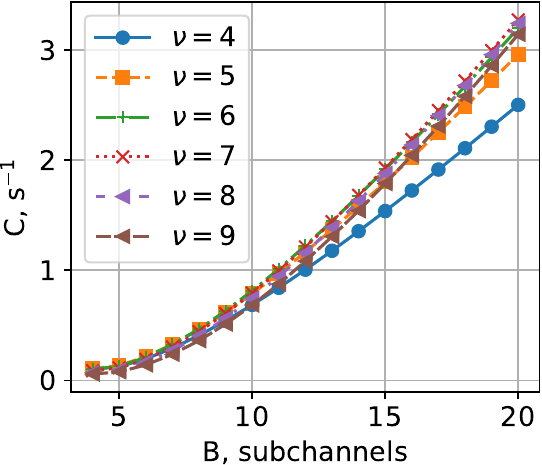}
		\caption{$PLR^{QoS} = 10^{-5}$.}
	\end{subfigure}
	\caption{Network capacity vs. bandwidth.}
	\label{fig:c_b_lax}
\end{figure}

Figure~\ref{fig:validation} shows PLR as a function of $\lambda$ for a given number of repetitions $\nu$ obtained with: (i) simulations (``Sim.'' markers), (ii) our analytical model (``An. model'' curves), and (iii) an analytical model of a similar LTE V2X channel access method~\cite{vinel2019cellularV2X} (``Model \cite{vinel2019cellularV2X}'' curves).
Although~\cite{vinel2019cellularV2X} does not consider DENM collisions, it assumes that both CAM and DENM packets are transmitted with distributed channel access.
So, we set the CAM size equal to the DENM size and the CAM periodicity equal to the average DENM generation time.
Figure~\ref{fig:validation} shows that our model has high accuracy for any $\nu$ because it considers repetitions as described in Section~\ref{sec:active-ue}.
In contrast, the accuracy of the model~\cite{vinel2019cellularV2X} significantly degrades for $\nu > 0$ because that model does not take into account the correlation of collisions between repetitions of various UEs when the delay budget $D^{QoS}$ is small, which results in PLR underestimation by an order of magnitude. 
Moreover, the model~\cite{vinel2019cellularV2X} significantly overestimates the network capacity.
If used to select transmission parameters in real systems, it will cause a violation of the reliability requirement.

Let us use the model to study how the network capacity depends on the bandwidth and the number of repetitions.
Figure~\ref{fig:c_b_lax} shows the results for two reliability requirements: (a) $PLR^{QoS} = 10^{-2}$ corresponding to the basic safety applications, (b) $PLR^{QoS} = 10^{-5}$ corresponding to the emerging V2X applications. 
We see that $\nu$ that maximizes the capacity mainly depends on $PLR^{QoS}$.
Three to four repetitions shall be used for $PLR^{QoS} = 10^{-2}$.
Note that the model~\cite{vinel2019cellularV2X} predicts $\nu = 6$ but setting such $\nu$ leads to 25\% lower capacity for $B = 10$ compared with the optimal $\nu$.
For $PLR^{QoS} = 10^{-5}$, six to seven repetitions are needed.
We observe that the curves for different $PLR^{QoS}$ have different shapes: (a) for $PLR^{QoS} = 10^{-2}$, we see a superlinear growth, (b) for $PLR^{QoS} = 10^{-5}$, the capacity is close to zero for $B < 6$ and then linearly grows, which aligns with \cite{karamyshev2020fast} and shows that in scenarios with low $D^{QoS}$ and $PLR^{QoS}$ the bandwidth below some threshold leads to almost zero capacity. 
Thus, the developed model can be used to select the optimal transmission parameters (e.g., the number of repetitions) and allocate the bandwidth to provide a given capacity.

\section{Conclusion}
\label{sec:conclusion}
In the letter, we have developed an analytical model of 5G V2X Mode 2 for sporadic traffic that allows estimating PLR and the network capacity for given scenario parameters.
In contrast to the existing models, our model takes into account the main peculiarities of Mode 2 and provides high accuracy at the PLR level of $10^{-5}$, which is the target for the emerging V2X scenarios.
The model allows finding the transmission parameters to maximize the network capacity or selecting the bandwidth for given QoS requirements and the traffic load.   

In future, we are going to extend our model to take into account the new features of Mode 2, e.g., resource reservation.

\bibliographystyle{IEEEtran}
\bibliography{biblio}

\end{document}